\documentclass[twocolumn,amsmath,amssymb,aps,pra,showpacs,tightenlines]{revtex4}
\usepackage{amsmath}
\usepackage{amssymb}
\usepackage{txfonts}
\usepackage{graphicx}
\usepackage{epstopdf}
\usepackage{dcolumn}
\usepackage{bm}
\usepackage[colorlinks=true,citecolor=blue,linkcolor=magenta]{hyperref}

\begin{document}
\title{Generation and amplification of high-order sideband induced by two-level atoms in a hybrid optomechanical system}
\author{Zeng-Xing Liu}
\author{Hao Xiong}\email{haoxiong1217@gmail.com}
\author{Ying Wu}
\affiliation{School of Physics, Huazhong University of Science and Technology, Wuhan, 430074, P. R. China}
\date{\today}

\begin{abstract}
It is quite important to enhance and control the optomechanically induced high-order sideband generation to achieve low-power optical comb and high-sensitivity sensing with an integratable structure. Here we present and analyze a proposal for enhancement and manipulation of optical nonlinearity and high-order sideband generation in a hybrid atom-cavity optomechanical system that is coherently driven by a bichromatic input field consisting of a control field and a probe field and works beyond the perturbative regime. Our numerical analysis with experimentally achievable parameters confirms that robust high-order sideband generation and typical spectral structures with non-perturbative features can be created even under weak driven fields. The dependence of the high-order sideband generation on the atomic parameters are also discussed in detail, including the decay rate of the atoms and the coupling parameter between the atoms and the cavity field. We show that the cutoff order as well as the amplitude of the higher order sidebands can be well tuned by the atomic coupling strength and the atomic decay rate. The proposed mechanism of enhancing optical nonlinearity is quite general and can be adopted to optomechanical systems with different types of cavity.
\end{abstract}
\pacs{03.65.Ta, 42.50.Wk}

\maketitle
\section{Introduction}
Cavity optomechanics which studies the interaction between light and mechanical oscillation \cite{Review1,Review2,Review3} has become a rapidly developing field recently and plays an important role in many fields of physics, including precision measurements \cite{precision measuement,precision measuement2, precision measuement1,Review16}, gravitational-wave detectors \cite{gravitational detection,gravitational detection0,gravitational detection1}, integrated optical components \cite{Optical polarizer}, cooling of mechanical oscillators \cite{groud state cooling,groud state cooling1}, and manipulation of light propagation \cite{nonreciprocity}. It has been shown that many effects observed in atomic systems can also occur in an optomechanical system through mechanical effects of light. A typical example is optomechanically induced transparency \cite{fast and slow} that the transmission of a probe field through an optomechanical device can be modulated all-optically using a strong control field in analogy to electromagnetically induced transparency. Other phenomena in atomic vapors, such as slow light and optical storase, have also been observed in optomechanical system \cite{omit1,omit10,omit2} due to the strong dispersion concomitant with the transparency window.

Combination of optomechanical systems with atoms gives rise to hybrid atom-cavity optomechanical systems, which provides the coherent manipulation of light with additional degrees of freedom and has undergone rapid progress in the past few years. This emerging subject leads to a variety of applications, including convenient control of optomechanical devices \cite{1}, single-photon transport \cite{2}, and resolving the vacuum fluctuations \cite{3}. Recently, it has been shown that fast and slow light effect can be enhanced and mutual switched in a hybrid atom-cavity optomechanical system \cite{fast light}, which may pave the way toward real applications, such as optomechanical memory and telecommunication \cite{memory}. All these effects involve the linearized description of optomechanical interactions, and can be well explained by the linearized Heisenberg-Langevin equations.

Optomechanical interaction is intrinsic nonlinear, and more and more attention has been dedicated to exploring the nonlinear property of optomechanical interaction in recent years \cite{Review4,hx1,hx2,s4}. Cavity optomechanics in the nonlinear regime has many interesting phenomena in both semiclassical and quantum mechanisms. In the semiclassical mechanism, optomechanical nonlinearity induced high-order sideband generation \cite{hsg,hsg2,hsg1,s3} and optomechanical chaos \cite{chaos,s1}. Similar to high-order harmonic generation in atomic systems \cite{harmonic}, the high-order sideband generation may find applications in many aspects, including generation of excellent optical-frequency combs \cite{three,s2} and precision measurement of DNA molecules properties \cite{DNA}. However, robust high-order sideband generation can only occur under ultrastrong driven fields in the nonperturbative regime \cite{hsg,hsg2}. In view of the potential applications of high-order sideband generation, an interesting question is concerned about that whether one can easily generate and amplify  the high-order sideband generation with experimentally system parameters. Furthermore, it is quite important to enhance and control the optomechanically induced high-order sideband generation to achieve low-power optical comb and high-sensitivity sensing with an integratable structure.

In this paper, we present and analyze a proposal for enhancement and manipulation of optical nonlinearity and high-order sideband generation in a hybrid atom-cavity optomechanical system that is coherently driven by a bichromatic input field consisting of a control field and a probe field, and works beyond the perturbative regime. We show that the two-level atoms can obviously strengthen the optomechanical nonlinearity in our scheme, and robust high-order sideband generation as well as typical spectral structures with non-perturbative features can be created even under weak driven fields. The features of high-order sideband generation can be easily adjust and amplify, which may be of great interest in view of its potential value on present-day photonic technology \cite{use1,use2}.

The paper is organized as follows. In Sec. \ref{sec:2}, we present the theoretical description of a hybrid optomechanical system and give the derivation of Heisenberg-Langevin equation of motion in the present of the two-level atoms. In Sec. \ref{sec:3}, we show that the two-level atoms play an important role in the hybrid optomechanical system which can greatly induced the generation and amplification of the high-order sidebands. Finally, a conclusion is summarized in Section \ref{sec:4}.

\section{physical Model and dynamical equation \label{sec:2}}

\begin{figure}[htb]
\centering\includegraphics [width=7.5cm] {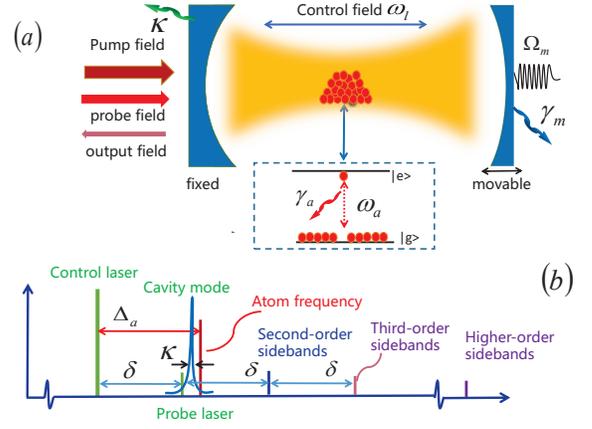}
\caption{(Color online) (a) The schematic diagram of our hybrid optomechanical system. An ensemble of $N$ identical two-level cold atoms, with transition frequency $\omega_{\rm{a}}$ and decay rate $\gamma_{\rm{a}}$, are embedded in the cavity field. Levels $|g\rangle$ and $|e\rangle$ denote, respectively, the ground and excited states of the atom. (b) Frequency spectrogram of a hybrid atom-cavity optomechanical system. The red line represents the atom frequency is detuned by $\Delta_{\rm{a}}$ from the control field. There are second-, third-, and high-order sidebands generation due to the nonlinear optomechanical interactions.}
\label{fig:1}
\end{figure}

A schematic description of a hybrid atom-cavity optomechanical system is shown in Figure. \ref{fig:1} (a). The system consists of a high-Q Fabry-Perot cavity, in which one mirror is movable and treated as a quantum mechanical harmonic oscillator with effective mass $m$ and angular frequency $\Omega_{\rm{m}}$, and an ensemble of $N$ identical two-level cold atoms imprisoned in a magneto-optical trap locating at the middle of the cavity field \cite{mode}. The mechanical oscillator and atoms are coupling to the cavity mode via optomechanical and dipolar interactions, respectively. In this letter, our system is driven by two fields: a strong control field with frequency $\omega_{\rm{l}}$ and a weak probe field with frequency $\omega_{\rm{p}}$. The Hamiltonian of our scheme can therefore be written as:
\begin{eqnarray}\label{equ:1}
\hat{H}_{\rm{0}} &=& \hbar\omega_{\rm{c}}\hat a^\dag\hat a+(\frac{\hat p^2}{2m}+\frac{1}{2}{m\Omega^2\hat x^2})+\hbar\sum_{\rm{i=1}}^{\rm{N}}\omega_{\rm{a}}\hat{\sigma}_{\rm{ee}}^{\rm{(i)}}\nonumber\\
&&+\hbar G\hat a^\dag\hat a\hat{x}+\hbar g\sum_{\rm{i=1}}^{\rm{N}}(\hat{a}\hat{\sigma}_{\rm{eg}}^{\rm{(i)}}+\hat{a}^{\dagger}\hat{\sigma}_{\rm{ge}}^{\rm{(i)}})+\hat{H}_{\rm{in}}
\end{eqnarray}
where the first term is the free Hamiltonian of the cavity field with annihilation and creation operators $\hat{a}$ and $\hat{a}^{\dagger}$. The second term denotes the energy of the mechanical oscillator with $\hat x$ and $\hat p$ being the position and momentum operators, respectively. The third term expresses the Hamiltonian of the two-level atom system, where $\hat{\sigma}_{\rm{eg}}^{\rm{(i)}}$ = $|e\rangle_{\rm{ii}}\langle g|$ for $e \neq g$ is the electronic projection operator and $e = g$ is the population operator of the $i$th atom, and $\omega_{\rm{a}}$ is the transition frequency between the ground state $|g\rangle$ and excited states $|e\rangle$. The fourth term describes the interaction of the cavity field and the movable mirror via radiation pressure with the optomechanical coupling strength $G$. The next term represents the coupling between the two-level atoms and cavity field, where $g = \mu\sqrt{\omega_{\rm{c}}/2\hbar V\varepsilon_{\rm{0}}}$ is the single-atom vacuum Rabi frequency with the dipole moment $\mu$, the cavity volume $V$, and the vacuum permittivity $\varepsilon_{\rm{0}}$. $\hat{H}_{\rm{in}} = \hat{H}_{\rm{control}} + \hat{H}_{\rm{probe}}$ where $\hat{H}_{\rm{control}} = i\hbar\varepsilon_{\rm{l}}(\hat{a}^{\dagger}e^{-i\omega_{\rm{l}}t} - \hat{a}e^{i\omega_{\rm{l}}t})$ and $\hat{H}_{\rm{probe}} = i\hbar(\hat{a}^{\dagger}\varepsilon_{\rm{p}}e^{-i\omega_{\rm{p}}t} - \rm{H.C}.)$.  $\varepsilon_{\rm{i}}=\sqrt {2\kappa P_{\rm{i}}/\hbar\omega_{\rm{i}}}\quad(i=l, p)$
are the amplitudes of the input field with $P_{\rm{l}}$ the power of the control field, and $P_{\rm{p}}$ the power of the prob field.
$\kappa$ is the total decay rate of the cavity field which contains an intrinsic loss rate $\kappa_{o}$ and an external loss rate $\kappa_{ex}$.

A previous study \cite{CO} has shown that the two-level atoms in a cavity field can be seen as a whole for the large atomic number $N$ in the case of most of the atoms are originally prepared in the ground state. We can therefore define:
\begin{eqnarray}
\hat{\rho}_{\rm{a}} &=& {\lim_{N \to \infty}}\sum_{\rm{i=1}}^{\rm{N}}(g_{\rm{i}}^{\ast}/g_{\rm{ac}})|1\rangle_{\rm{ii}}\langle2|,
\end{eqnarray}
as a collective transition operator satisfying the bosonic commutation relation $[\hat{\rho}_{\rm{a}}, \hat{\rho}_{\rm{a}}^{\dagger}] = 1$. $g_{\rm{ac}}=g\sqrt{N} = \sqrt{\sum_{\rm{i=1}}^{\rm{N}}|g_{\rm{i}}|^2}$ is the collective vacuum Rabi frequency equaling to the total coupling strength between the atomic en bloc and the cavity field, which will enhance with the increasing of atomic number $N$. Transforming the Hamiltonian into the rotating frame at the frequency $\omega_{\rm{l}}$ based on $\hat{H}_{\rm{l}} = \hbar\omega_{\rm{l}}\hat{a}^{\dagger}\hat{a}$ and
$U_{\rm{t}} = e^{-i\hat{H}_{\rm{l}}t/\hbar} = e^{-i\omega_{\rm{l}}\hat{a}^{\dagger}\hat{a}t}$ gives the following Heisenberg equations:
\begin{eqnarray}\label{equ:2}
\dot{\hat x} &=& \frac{\hat{p}}{m},\nonumber\\
\dot{\hat p} &=& -m\Omega_m^2\hat x-\hbar G\hat a^\dag \hat a-\gamma_m\hat p+\sqrt{2\gamma_{m}}\hat F_{th}(t),\nonumber\\
\dot{\hat a} &=& -(2\kappa+i\Delta_{c}+iG\hat x)\hat a-ig_{\rm{ac}}\hat\rho_{\rm{a}}\nonumber \\
  &&+\varepsilon_{\rm{l}}+\varepsilon_{\rm{p}}e^{-i\delta t}+\sqrt{2\kappa}\hat a_{in}(t),\nonumber \\
\dot{\hat\rho}_{\rm{a}} &=& -(\gamma_{a}+i\Delta_{a})\hat\rho_{\rm{a}}-ig_{\rm{ac}}\hat{a}+\sqrt{2\gamma_{a}}\hat \rho_{th}(t),
\end{eqnarray}
where $\Delta_{c}$ = $\omega_{c}-\omega_{l}$, $\Delta_{a}$ = $\omega_{a}-\omega_{l}$, and $\delta$ = $\omega_{p}-\omega_{l}$ are the detunings of the cavity field frequency, the atomic transition frequency, and the probe field frequency, with the control field frequency, respectively. Here, we have classically introduced the damping $\gamma_{m}$ of the mirror and the decay rate $\gamma_{a}$ of the two-level atom. The thermal Langevin force $\hat F_{th}(t)$, resulting from the coupling of the mechanical oscillator to the environment, satisfy the correlation function $\langle{\hat F_{th}(t)\hat F^{\dagger}_{th}(t^{'})}\rangle= \gamma_{m}\int e^{-i\omega(t-t^{'})}[\rm{coth}(\hbar\omega/2k_{\rm{B}}T)+1]d\omega/2\pi\Omega_{m}$, where $k_{\rm{B}}$ and $T$ are the Boltzmann constant and the temperature of the reservoir of the mechanical oscillator, and the mean value $\langle{\hat F_{th}(t)}\rangle = 0$ \cite{noise}.
 The quantum noise of the optical cavity and the two-level atoms are represented by $\hat a_{th}(t)$ and $\hat \rho_{th}(t)$. Without loss of generality, $\hat a_{th}(t)$ and $\hat \rho_{th}(t)$ resulting from the environment, obey the correlation function $\langle{\hat a_{th}(t)}{\hat a^{\dagger}_{th}(t^{'})}\rangle = \delta(t-t^{'})$, $\langle{\hat \rho_{th}(t)}{\hat \rho^{\dagger}_{th}(t^{'})}\rangle = \delta(t-t^{'})$, and the average value $\langle{\hat a_{th}(t)}\rangle = 0$, $\langle{\hat\rho_{th}(t)}\rangle = 0$ \cite{noise}, which is demonstrated to be valid in the concerned weak-coupling regime \cite{Review1}. In this work, we are interested in the average response of the system, so the operators can be reduced to their expectation values, i.e., $x(t)\equiv\langle{\hat x(t)}\rangle$, $p(t)\equiv\langle{\hat p(t)}\rangle$, $a(t)\equiv\langle{\hat a(t)}\rangle$, and $\rho_{\rm{a}}(t)\equiv\langle{\hat \rho_{\rm{a}}(t)}\rangle$. Using a so-called mean-field approximation, viz., $\langle{xa}\rangle=\langle{x}\rangle\langle{a}\rangle$, the Heisenberg-Langevin equations can be written as:
\begin{eqnarray}\label{equ:3}
\frac{d\beta}{dt} &=& M\beta + \xi
\end{eqnarray}
where $\beta = (p, x, a, \rho_{\rm{a}})^{\rm{T}}$, $\xi = (0, 0, \varepsilon_{\rm{l}}+\varepsilon_{\rm{p}}e^{-i\delta t}, 0)^{\rm{T}}$ and
\begin{eqnarray}
M = \left(
  \begin{array}{cccc}
    -\gamma_{\rm{m}} & -m\Omega_{\rm{m}}^{2} & -\hbar Ga^{\dagger} & 0 \\
    1/m & 0 & 0 & 0 \\
    0 & 0 & \Re & -ig_{\rm{ac}} \\
    0 & 0 & -ig_{\rm{ac}} & -(\gamma_{\rm{a}}+i\Delta_{\rm{a}}) \\
  \end{array}
\right)\nonumber
\end{eqnarray}
where $\Re = -(2\kappa+i\Delta_{\rm{c}}+iGx)$.
This set of equations are describe the evolution of the hybrid optomechanical system. Note that, if $g_{\rm{ac}} = 0$, our system is reduced to a traditional optomechanical system, where optical polarizer \cite{Optical polarizer}, optical transmission \cite{nonreciprocity} and fiber-optical frequency comb \cite{three} have been discussed in the perturbative regime. In our study, we numerically analyze the nature of the optomechanical interaction beyond the conventional linearized description, and give a full presentation of the non-perturbative effects in our scheme.

\section{Results and discussion \label{sec:3}}

The Heisenberg-Langevin equations discussed above are nonlinear, and it is very difficult to give an analytical solution. A procedure that ignores these nonlinear terms has been generally adopted in many previous works \cite{precision measuement,precision measuement2,fast and slow}. Using the linearization of the Heisenberg-Langevin equations, optomechanically induced transparency in a hybrid atom-cavity optomechanical system has been discussed in a previous work \cite{fast light}, and it is shown that the fast and slow light can be enhanced and mutual switched \cite{CO}.
Following the analytical perturbation method of describing the linearized optomchanical interaction, each operator average value can be written as a sum of its steady-state value and fluctuations from the stationary state, viz., $ a= \bar{a}+\delta a$ and $ x= \bar{x}+\delta x$. In the case that the control field is much stronger than the probe field, we can obtain the steady-state solution of Eqs. (\ref{equ:3}) \cite{fast and slow}:
\begin{eqnarray}
  \bar{a} &=& \frac{\varepsilon_{\rm{l}}}{2\kappa+i\Delta_{\rm{c}}+\Theta},\qquad\bar{x}=\frac{\hbar G|\bar{a}|^2}{m\Omega_{\rm{m}}^{2}}
\end{eqnarray}
\begin{figure}[htb]
\centering\includegraphics[width=7.5cm] {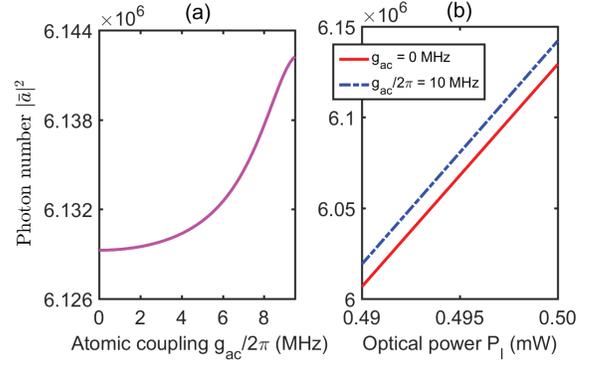}
\caption{(color online) The photon number $|\bar{a}|^2$ as a function of (a) the atomic coupling strength $g_{\rm{ac}}$, and (b) the power of the control field $P_{\rm{l}}$. The other parameters are $m$ = 10 $\rm{ng}$, $\Omega_{\rm{m}}/2\pi$ = 10 $\rm{MHz}$, $G/2\pi$ = 0.4 $\rm{GHz/nm}$, $\kappa/2\pi$ = 1 $\rm{MHz}$, $\gamma_{m}/2\pi$ = 100 $\rm{Hz}$, $\gamma_{a}/2\pi$ = 2.875 $\rm{MHz}$, $\Delta_{c} = \Delta_{a}= \Omega_{\rm{m}}$, wavelength of the laser $\lambda$ = 2$\pi c/\omega_{\rm{l}}$ = 794.98 $\rm{nm}$, $P_{\rm{l}}$ = 0.5 $mW$, and $\varepsilon_{\rm{l}}$=  0.05$\varepsilon_{\rm{p}}$. All the parameters are chosen based on the recent experiment \cite{parameter1,parameter2} and can be achieved under current technology.}
\label{fig:2}
\end{figure}
where $\Theta = iGx+g_{\rm{ac}}^{2}/(\gamma_{\rm{a}}+i\Delta_{\rm{a}})$. We can see that the intracavity photon number $|\bar{a}|^2$ shows a strong dependence on the two-level atoms. Figure. \ref{fig:2} (a) plots the photon number $|\bar{a}|^2$ varies with the total coupling strength $g_{\rm{ac}}$. It is shown that the photon number rapidly increases with the atomic coupling strength within the regime
$g_{\rm{ac}}/2\pi\in[0, 10]$ $\rm{MHz}$, which reminds us of the possibility of enhancing the optomechanical nonlinearity in our system by introducing the atoms. In addition, the relationship between the photon number $|\bar{a}|^2$ and the control power was plotted in Fig. \ref{fig:2} (b). It can be found that the number of photons increases linearly with the control power. Comparing with the bare cavity (as red line shows), the photon number is obviously enhanced in the presence of the two-level atoms (as blue dash-dotted line shows).

Some interesting phenomena appear when we consider the nonlinear terms in Eqs. (\ref{equ:3}), such as higher-order sidebands generation \cite{hsg1}, optomechanical induced sum (difference)-sideband generation \cite{difference}, and Kuznetsov-Ma solitons in optomechanical system \cite{Soliton}. In the present work, the nonlinear terms in a hybrid atom-cavity optomechanical system is more complex due to the coupling between the cavity and the two-level atoms, and we mainly discuss the effect of the atoms on high-order sideband generation in our scheme. Eqs. (\ref{equ:3}) are ordinary differential equations, and we use Runge-Kutta method to solve these equations. We use $a|_{t=0} = 0$, $x|_{t=0} = 0$, $p|_{t=0} = 0$, and $\rho_{a}|_{t=0} = 0$ as the initial condition \cite{Runge}. The output field $s_{\rm{out}}(t)$ can be obtained by using the standard input-output notation $s_{\rm{out}}(t)=s_{\rm{in}}(t)-\sqrt{2\kappa}a(t)$. $s_{\rm{in}}(t)$ = $\varepsilon_{\rm{l}}e^{-i\omega_{\rm{l}}t} + \varepsilon_{\rm{p}}e^{-i\omega_{\rm{p}}t}$ is the driving field containing a control field and a probe field, and the amplitudes of the control field $\varepsilon_{\rm{l}}$ and the probe field $\varepsilon_{\rm{p}}$ are normalized to a photon flux at the input of the cavity \cite{hsg2}. The output spectrum can be obtained by performing fast Fourier transform of $s_{out}(t)$, i.e., $S(\omega)\propto | \int_{-\infty}^{\infty} s_{out}(t)e^{-i\omega t}dt |$ with $\omega$ is the spectrometer frequency.
\begin{figure}[htb]
\centering\includegraphics [width=7.5cm] {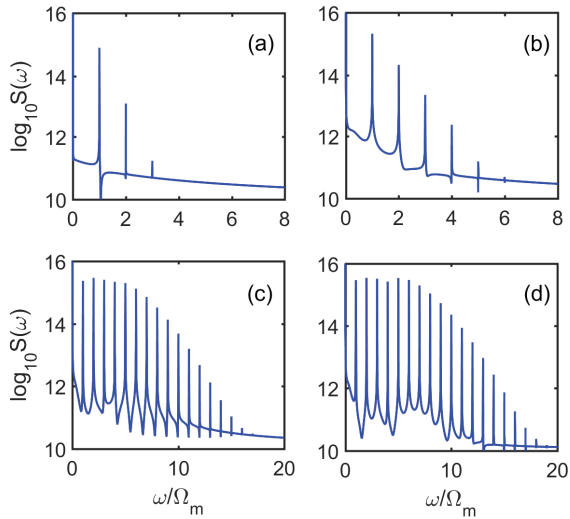}
\caption{(Color online) The higher-order sidebands generation spectra output from the hybrid optomechanical system are shown with different atomic coupling parameter $ g_{ac}$: (a)$g_{ac}/2\pi$ = 0 $\rm{MHz}$, (b) $g_{ac}/2\pi$ = 12.6 $\rm{MHz}$, (c) $g_{ac}/2\pi$ = 12.9 $\rm{MHz}$, (d) $g_{ac}/2\pi$ = 13.2 $\rm{MHz}$, respectively. The other parameters for simulation are chosen as the same as in Fig. \ref{fig:2}.}
\label{fig:3}
\end{figure}
The physical picture of this transform is that the output field of the system can be formally expressed in the time domain by $s_{\rm{out}}(t) = \sum_{i=0}^{n}A_{i}e^{-i(\omega_{c}\pm \delta)t}$ $(i = 0, 1, 2\cdot\cdot\cdot)$, where $A_{i}$ is the $i$th transmission coefficient of the output field, and we can see that there are frequencies $\omega_{c}\pm \delta$ generating in the frequency spectrum of the output field. For simplicity, we change the time evolution of the output field from the time domain to the frequency domain by doing fast Fourier transform, which can be directly measured \cite{Fourier}.
We use experimentally realizable parametric values of the optomechanical system \cite{parameter1,parameter2} in our numerical calculation. Before passing to the results of the numerical calculation, we need to emphasize that the spectra obtained shift a frequency $\omega_{\rm{l}}$, because the Heisenberg-Langevin equations describe the time evolution of the optical field in a frame rotating at the control frequency $\omega_{\rm{l}}$.

To illustrate the remarkable influence of the two-level atoms on high-order sideband generation,
the frequency spectra output from the hybrid optomechanical system is shown in Fig. \ref{fig:3}. We can clearly see that the cutoff order and the amplitude of higher-order sidebands can be significantly enhanced by two-level atoms. Under the weak driving field, the power of the control field $P_{\rm{l}}$ = 0.5 $\rm{mW}$ and probe field $P_{\rm{p}}$ = 1.25 $\mu W$, and the atomic coupling strength is switched off, i.e., $g_{ac} = 0$, as shown in Fig. \ref{fig:3} (a), there are only the first-order, second-order and third-order sidebands obviously appear in the frequency spectra. Furthermore, on the whole the higher the order we considered, the small the amplitude obtained, which clearly works in the perturbative regime.
\begin{figure}[htb]
\centering\includegraphics [width=7.5cm] {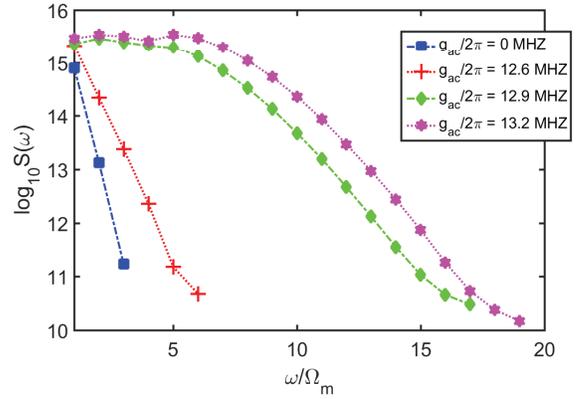}
\caption{(Color online) The order and amplitudes of the higher-order sidebands under different atomic coupling strength $g_{ac}$. The other parameters for simulation are chosen as the same as in Fig. \ref{fig:2}.}
\label{fig:4}
\end{figure}
Advantageously, we find that resonantly enhanced feedback-backaction arising from nonlinear optomechanical interaction can be substantively reinforced in the presence of two-level atoms. Different from Fig. \ref{fig:3} (a), due to the reinforcement of optomechanical nonlinearity between cavity field and mechanical oscillator induced by two-level atoms, not only the cutoff order but the amplitude of the higher-order sidebands are enhanced as well. As shown in Fig. \ref{fig:3} (b), the value of the atomic coupling strength $g_{ac}/2\pi$ = 12.6 $\rm{MHz}$ and $\Delta_{c}=\Delta_{a}$, viz., $\omega_{c}=\omega_{a}$, namely, the cavity field frequency is resonant with the atomic transition frequency, the sideband spectrum ends up at the order of 6 and the intensity of each sideband is strengthened. From above discussion, we can note that the two-level atoms play a vital role in an optomechanical system, especially in the generation and amplification of higher-order sidebands, and we call it atomic-resonant-induced high-order sideband generation.

To further enhance optical nonlinearity and high-order sideband generation, we continue to increase the atomic coupling strength $g_{ac}/2\pi$ = 12.9 $\rm{MHz}$ and the frequency spectrum output from the hybrid optomechanical system is shown in Fig. \ref{fig:3} (c). The frequency spectrum is quite different with respect to Fig. \ref{fig:3}(b), for example, the amplitude of the second-order sideband is larger than the first-order sideband. A plateau appears in the sideband spectrum where all the sidebands have nearly the same strength, which is the typical nonperturbative feature of the high-order sideband generation \cite{hsg}.
We can clearly see that the order of the sideband in frequency spectrum visibly ends up at the order of 16, and the intensity of each sideband is also distinctly enhanced.
To get a robust high-order sideband generation, we use an even stronger atomic coupling strength $g_{ac}/2\pi$ = 13.2 $\rm{MHz}$, and the frequency spectrum of the output field is shown in Fig. \ref{fig:3} (d), in which the cutoff order is extended to about 19 and the strength of the sidebands are further enhanced. We note that, especially, the strength of the high-order sidebands can be easily modulated and amplified by adjusting the atomic coupling strength, which is more superior and achievable than the power-driven-induced high-order sideband generation discussed previously \cite{hsg2}.
Such a prominent atomic-resonance-induced high-order sideband generation in view of our scheme reminds us of the possibility of realizing adjustable optical-frequency-comb with weak power.

\begin{figure}[htb]
\centering\includegraphics[width=8.5cm] {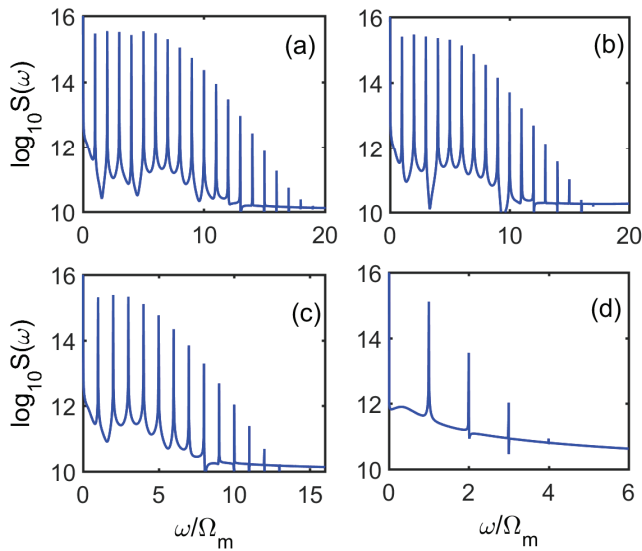}
\caption{(Color online) The higher-order sidebands generation spectra output from the hybrid optomechanical system are shown under different atom decay rate $\gamma_{a}$: (a) $\gamma_{a}/2\pi$ = 2.875 $\rm{MHz}$, (b) $\gamma_{a}/2\pi$ = 3.306 $\rm{MHz}$, (c) $\gamma_{a}/2\pi$ = 3.738 $\rm{MHz}$, (d) $\gamma_{a}/2\pi$ = 4.169 $\rm{MHz}$, respectively. The atomic coupling strength $g_{ac}/2\pi$ = 13.2 $\rm{MHz}$ and the other parameters are the same in Fig. \ref{fig:2}.}
\label{fig:5}
\end{figure}
We observe a high dependence of the high-order sideband generation on the two-level atoms and more clear results are shown in Fig. \ref{fig:4} with the same experimental parameters. A previous study \cite{three} has shown that the coupling between the atoms and optical cavity have three different forms, i.e., the under-coupling, critical-coupling and over-coupling. It can be found that when the atomic coupling in the under-coupling regime, the order as well as intensity of the sidebands have not been greatly enhanced,
and a typical nonperturbative feature of the high-order sideband generation does not appears (as the green dash-dotted line and red dotted line shown).
Different from the under-coupling case, both the order and the amplitudes of the high-order sidebands have been reinforced, and a typical nonperturbative feature of the high-order sideband generation appears in the output spectrum when the coupling strength reaches the critical value (as the green dash-dotted line shown). However, when the atomic coupling exceeds the critical value into the over-coupling range, the growth trend of the sideband intensity and order is obviously slowed down (as the pink dotted line shown). The atomic coupling $g_{\rm{ac}}$ is proportional to the atom number $N$, and the photon number in the cavity field will tend to saturation when the atom number reaches a certain critical value. Namely, the process of the photon annihilation and phonon creation will tend to be stable.

In what follows, we discuss the effects of the atomic attenuation on high-order sideband generation. We plot the frequency spectra output from the hybrid optomechanical system under different atomic decay rate $\gamma_{\rm{a}}$ in Fig. \ref{fig:5}. Evidently, we can see that the order as well as amplitude of the high-order sidebands are closely related to the damping of the two-level atoms. For the case that the atom decay rate $\gamma_{a}/2\pi$ = 2.875$\rm{MHz}$, a robust high-order sideband generation appears in the frequency spectrum of the output field, as shown in Fig. \ref{fig:5} (a) (same as Fig. \ref{fig:4} (d)). As expected, the order and intensity of the sidebands is decreased when we reinforce the decay rate of the atoms. The results are shown in Fig. \ref{fig:5} (b) and (c), the intensity of the sidebands are obviously weakened and the cutoff order of the high-order sidebands are reduced to 15 and 12 corresponding to the atom decay rate $\gamma_{a}/2\pi$ = 3.306 $\rm{MHz}$ and $\gamma_{a}/2\pi$ = 3.738 $\rm{MHz}$, respectively. A remarkable result is shown in Fig. \ref{fig:5} (d) when we again increase the atomic damping $\gamma_{a}/2\pi$ = 4.169 $\rm{MHz}$. We can see that the order of the sidebands are drastically reduced, and there are only the first-order, second-order and third-order sidebands visible in transmission spectrum and the amplitude of each sideband is greatly weakened. The physical picture of such result is that with the increase of atomic attenuation rate, the effective transition between the atomic energy levels will be impeded. Namely, the atoms in high energy state will soon decay to low energy state which mean that the optomechanical nonlinearity can not be effectively improved and, therefore, the effects of atomic-resonant-induced high-order sideband generation are also suppressed.

In addition, it is not difficult to find that the linewidth of the high-order sidebands spectra are very narrow and almost to a line, which can be well explained by the uncertainty relations of time and frequency. According to the time-frequency uncertainty relation $\Delta\omega\Delta t\sim2\pi$, we can calculate the uncertainty of the frequency as $\Delta\omega\sim{2\pi/\Delta t}$. In this woke, we use two continuous-wave laser to drive the hybrid  optomechanical system and it lasts about infinity, i.e., $\Delta t\rightarrow+\infty$. Obviously, the uncertainty of the frequency $\Delta\omega\rightarrow0$. In this way, we can well explain the relatively narrow of the high-order sidebands generation.

\section{Conclusion \label{sec:4}}

In summary, we analyze high-order sideband generation in a hybrid atom-cavity optomechanical system, and we show that the two-level atoms embedded in the cavity field can dramatically enhance optomechanical nonlinearity, and the photon number of the optical cavity and the displacement of the mechanical oscillator are reinforced. We find that a typical spectral structure with non-perturbative features can be created even under weak driven fields due to the atomic-resonant-induced high-order sideband generation.
The dependence of the high-order sideband generation on the atomic parameters are discussed in detail, including the atomic coupling strength and atomic decay rate. We reveal that the cutoff order as well as amplitude of the sidebands can be well tuned by the atomic attenuation rate and coupling  strength between the atom and the cavity field. This investigation may provide further insight into the understanding of the nonlinear optomechanical interactions beyond the perturbative regime and find applications in optical comb with weak power.

\begin{acknowledgments}
This work is supported by the National Basic Research Program of China (Grant No. 2016YFA0301203) and NSF of
China (Grant Nos. 11405061, 11375067, and 11574104).
\end{acknowledgments}

\end{document}